\documentclass[preprintnumbers,amsmath,amssymb,superscriptaddress,twocolumn,showpacs]{revtex4}
\usepackage{graphicx}
\usepackage{dcolumn}
\usepackage{bm}

\begin{document}

\title{QED with a spherical mirror}
\
\author{G. H\'etet}
\affiliation{
Institut for Experimental physics, University of Innsbruck, Technikerstrasse 25, A-6020 Innsbruck, Austria}
\affiliation{
Institut for Quantum-optics and Quantum information, Austrian Academie of science, Otto-Hittmair-Platz 1, A-6020 Innsbruck, Austria}

\author{L. Slodi\v{c}ka}
\affiliation{
Institut for Experimental physics, University of Innsbruck, Technikerstrasse 25, A-6020 Innsbruck, Austria}

\author{A. Glaetzle}
\affiliation{
Institut for Quantum-optics and Quantum information, Austrian Academie of science, Otto-Hittmair-Platz 1, A-6020 Innsbruck, Austria}

\author{M. Hennrich}
\affiliation{
Institut for Experimental physics, University of Innsbruck, Technikerstrasse 25, A-6020 Innsbruck, Austria}

\author{R. Blatt}
\affiliation{
Institut for Experimental physics, University of Innsbruck, Technikerstrasse 25, A-6020 Innsbruck, Austria}
\affiliation{
Institut for Quantum-optics and Quantum information, Austrian Academie of science, Otto-Hittmair-Platz 1, A-6020 Innsbruck, Austria}


\begin{abstract}
We investigate the Quantum-Electro-Dynamic properties of an atomic electron close to the focus of a spherical mirror. We first show that the spontaneous emission and excited state level shift of the atom can be fully suppressed with mirror-atom distances of many wavelengths. A three-dimensional theory predicts that the spectral density of vacuum fluctuations can indeed vanish within a volume $\lambda^3$ around the atom, with the use of a far distant mirror covering only half of the atomic emission solid angle.
The modification of these QED atomic properties is also computed as a function of the mirror size and large effects are found for only moderate numerical apertures. We also evaluate the long distance ground state energy shift (Casimir-Polder shift) and find that it scales as $(\lambda/R)^2$ at the focus of a hemi-spherical mirror of radius $R$, as opposed to the well known $(\lambda/R)^4$ scaling law for an atom at a distance $R$ from an infinite plane mirror. Our results are relevant for investigations of QED effects, and also free space coupling to single atoms using high-numerical aperture lenses.
\end{abstract}
\maketitle

Spontaneous emission and level shifts of atoms can be notably altered by placing them between mirrors. By modifying the electromagnetic mode structure interacting with the atomic electron \cite{Pur46}, one obtains a significant change in these quantum-electrodynamic (QED) atomic properties. Most experimental studies make use of high finesse cavities \cite{Bru94, Hoo00, Pin00, Hei87,Kre04,Suk92} to see the effects. Another way to change the properties of single emitters is to place other identical atoms close-by as originally propounded by Dicke \cite{Dic54}. To observe large QED effects in this case, the dipole emission patterns have to overlap, which requires the atoms to be very close to each other. Such effects were analyzed using two trapped ions \cite{DVo96}, but the Coulomb interaction between the ions restricted their distance to a few microns. The interaction between two neutral atoms is however not overwhelmed by the Coulomb force. Using the large dipole moments of nearby Rydberg atoms localised in a dipole trap, entanglement between neutral atoms was recently demonstrated \cite{Wil10,Ise10}.

In general, an atom close to a single mirror already provides a very efficient way to investigate QED effects. The resonance fluorescence of a Doppler cooled Barium ion was reflected back onto itself in \cite{Esc01}, using a large numerical aperture (NA) lens and a mirror that was 30 cm away from the ion. In this experiment, the description of the interaction between the atom and the modified electromagnetic field, or the mirror image, is very similar to the direct dipole-dipole coupling between two real atoms.
Here, due to the high numerical aperture of the collection lens, the mode structure was altered significantly even if the mirror was many wavelengths away from the ion. A 1\% change in the decay rate was measured and found to be mostly limited by the collection solid angle and residual atomic motion. Such a system also leads to a vacuum-induced level shift in a laser-excited atom. This has been measured in \cite{Wil04} and found in good agreement with theoretical work \cite{Dorner}.

A closely related field of research investigates the absorption of photons from single atoms in free space. Theory predicts that the best possible absorption between an incoming field and single atom arises when the incoming field matches the spatial atomic radiation mode \cite{Sto09,Lin07}. Recent experiments have demonstrated substantial extinctions from single molecules \cite{Zum08,Ger07,Wri08}, atoms \cite{Tey08,Slo10} and quantum dots \cite{Vam07} in free space, thus showing the potential of free space coupling with high NA optics for fundamental investigation of light-matter interactions.

The above mentioned studies make use of the interaction of real photon with single atoms. There has also been a rapid increase in the number of experiments related to Casimir forces between dielectric materials, which are the result of the modification of the mode density of virtual photons. Such studies are now being undertaken with an unprecedent level of precision (see for example \cite{Mun09} and references therein). The comparison with the theory in this field is now reaching good agreement for some geometries, and over a wide range of materials.  The possibility to use these measurements to gain a better control over nano-mechanical systems, and for precise tests of QED has been a major force driving this research. Although many geometries have been investigated theoretically over the past decades \cite{Zah10}, there are still many investigations relating to the sign of the force \cite{Ken06}, or the accuracy of the proximity force approximations \cite{Rod06}, for estimating Casimir shifts of various materials.

For atoms close to dielectrics, the modification of the ground state level shift (Lamb shift) yields the well known Casimir-Polder force \cite{Cas48}, that was observed in \cite{Suk92} for a plane mirror geometry.  The Casimir-Polder force was not reported nor calculated for single well localized atoms around complex opened 3D geometries. It is expected that such investigations would also provide efficient ways to test the behavior of vacuum fluctuations.

Here, we demonstrate that a spherical mirror covering half of an atomic dipole emission profile can fully suppress its spontaneous emission and excited state level shifts provided the mirror is close enough to allow temporal interference of the field amplitudes (Markov limit).
We also calculate the shift of the ground state, the Lamb shift, as a function of mirror distance and found a scaling law that is more favorable than the plane mirror geometry for observing large shifts. Due to the development in the control of atomic motion \cite{Bla08} and mirror and lens fabrication \cite{Sor07,Str10,Shu10}, these effects are now within experimental reach.

The modification of spontaneous emission and level shifts is first calculated using a one-dimensional model where the electron radiates along the mirror axis. We find full suppression of the vacuum fluctuations coupling to the atomic electron, even with a single mirror.
The physical origin of this complete cancelation lies in the high spatial and temporal interference between the plane waves modes going to the mirror and the modes going to free space so that the density of vacuum fluctuations can reach zero around this idealized atom. Similar calculations were performed by several authors to model more realistic scenarios (see for example \cite{Dorner,Mil94} and references therein) with the inclusion of free
space vacuum modes that do not interfere with the mirror modes; therefore full extinction of spontaneous mission was not explicitly considered.
We furthermore extend the one-dimensional calculations to a three-dimensional theory that goes beyond the paraxial approximation, and demonstrate that the effective coupling to vacuum modes around an atomic electron can also reach zero within a volume $\lambda^3$ around the focus of a spherical mirror.
In the spherical basis, such an effect can be understood as the result of an interference between even and odd spherical modes. Last, the long-distance
ground-state energy shift (Casimir-Polder shift) is evaluated using the complete set of normal modes of the spherical mirror.
We find that the Casimir-Polder shift scales as $(\lambda/R)^2$ for a half mirror of radius $R$ and atomic transition wavelength $\lambda$, as opposed to the well known $(\lambda/R)^4$ scaling law for a plane mirror, where $R$ is the mirror distance from the atom.

There exists a wealth of studies on this topic. We would like to point out that, to the best of our knowledge, theoretical efforts concentrated on plane geometries \cite{Mil94, Kas05, Kak93, Bar74}, on cavities in the paraxial approximation \cite{Pur46,Hei87}, dielectric spheres \cite{Duc97,Che87,Hag00,Mil97,Boy68}, or parabolic mirrors \cite{Stobinska}.
Closely related work was however done for a large spherical open-cavity by J.M. Daul and P. Grangier \cite{Daul}. Strong enhancement and inhibition of vacuum fluctuations were found with moderate cavity finesses and a full set of normal modes were derived for the symmetric geometry. The case of a single mirror was found by setting the second mirror reflectivity to zero in a more general formula for the density of vacuum fluctuations in an asymmetric cavity. In section \ref{sec2}, we present a different route towards finding the normal mode amplitudes in the single mirror geometry, that we further use for the Casimir-Polder shift calculation.

We would like to stress that all the calculations are performed in the limit where the mirror is many wavelengths away from the atom ($k_0 R\gg 1$).
This simplifies the theoretical treatments, and experimental approach greatly, yet allowing strong QED effects to be observed. Let us also mention here also that, since we use a single mirror, the QED effects for an exited atom are here easier to understand as a modulation of the electron coupling to certain modes rather than a change in the mode density \cite{Hei87}.



\section{Atomic decay and level shifts}

\begin{figure}[!h]
\includegraphics[width=8cm]{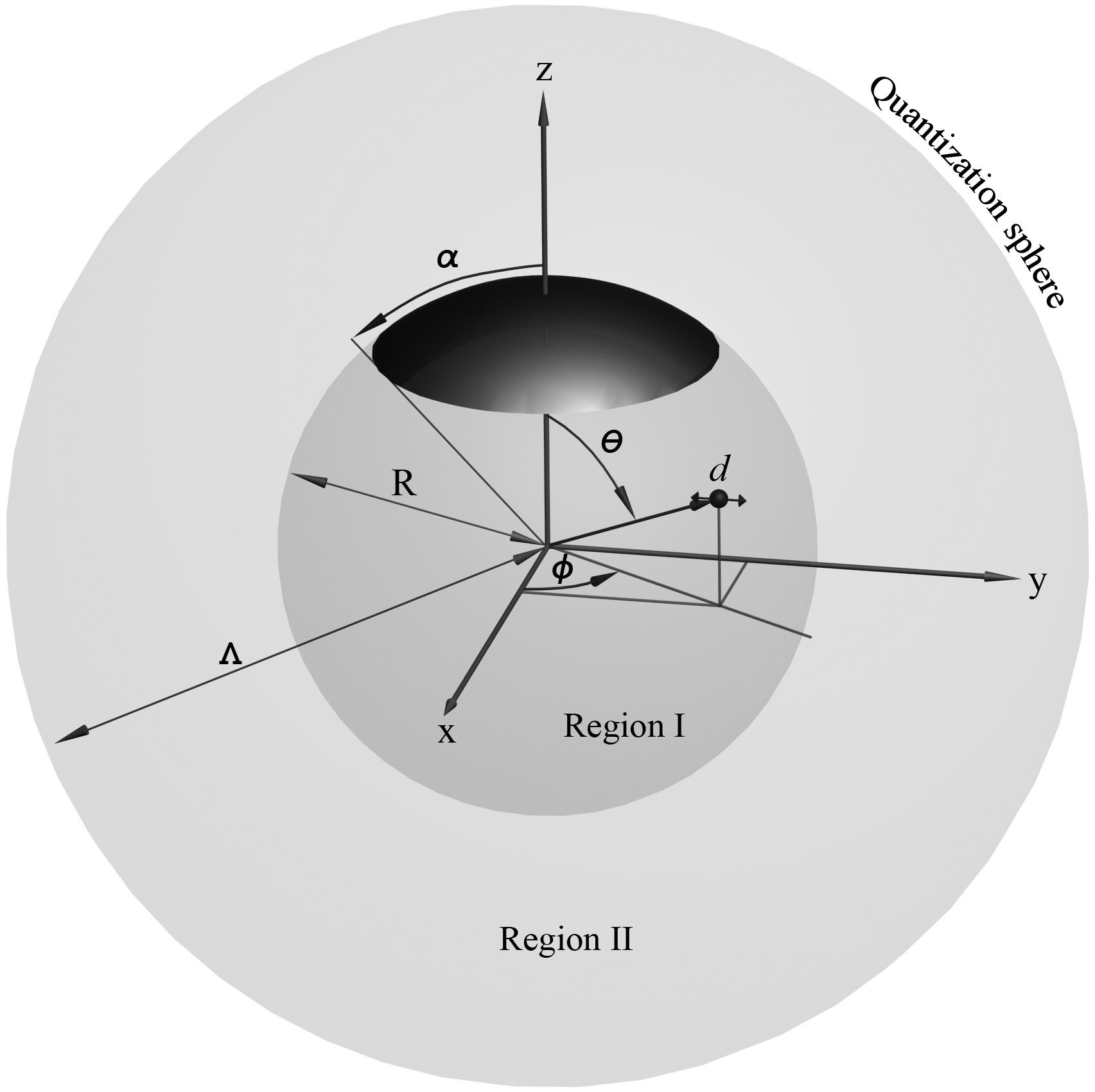}
\caption{Schematics of the mirror, and notations used to calculate the QED properties for an atomic dipole $\vec{d}$ in region I. The mirror has an aperture $\alpha$ and a radius of curvature $R$.}
\label{notations}
\end{figure}
It is well known that coupling an initially excited atom to a reservoir of electromagnetic field modes in the vacuum state yields a spontaneous decay to the ground state and shifts its excited state energy. Also, when the atom is in the ground state, its energy is shifted due to absorption and re-emission of virtual photons, the so-called Lamb shift. When the electromagnetic field modes are modified by nearby dielectric boundaries, these QED properties are therefore also changed.
Another picture, using radiation self-reaction only, can also be employed to describe the modification of QED properties \cite{Mil94}.
To find the relative contribution of both vacuum and self-reaction mechanisms, the dynamics of the corresponding quantities in the differential equation one wishes to interpret, has to be Hermitian \cite{DDC,Mes90}. When this is done, both effects are found to contribute to the same amount.

The free part of the light-atom Hamiltonian is the sum of the atomic and optical rest energies
\begin{eqnarray}
\hat{H}_0= \frac{1}{2}\hbar\omega_{0}\hat{\sigma}_z + \sum_\mu \hbar\omega_{{\mu}}\Big[\hat{a}_{{\mu}}^{\dagger}\hat{a}_{{\mu}}+\frac{1}{2}\Big],
\end{eqnarray}
where $\hat{\sigma}_z=\hat{\sigma}_{11}-\hat{\sigma}_{22}$ is the population difference between the upper and lower atomic states, and $\hat{a}_{{\mu}}$ the creation operator for a photon in a mode $\mu$ of the reservoir.
$\omega_0$ is the atomic transition frequency, $\omega_\mu$ the frequency of the optical mode $\mu$. The frequency of the optical modes is quantized using the boundary condition on a large cavity. In spherical coordinates we use a large quantization sphere of radius $\Lambda$, as depicted in Fig.~\ref{notations}.

The interaction Hamiltonian in the Coulomb gauge and in the electric dipole approximation is
\begin{eqnarray}
\hat{H}_{\rm int}=-\frac{e}{m c}\hat{A}(\vec{r},t)\cdot\hat{p}+\frac{e^2}{2mc^2}\hat{A}^2(\vec{r},t),
\end{eqnarray}
where $\hat{p}$ is the momentum of the atomic electron, $\vec{r}$ its position (i.e the position of the atomic nucleus in the electric dipole approximation) and $m$ its mass. $\hat{p}$ will be written in terms of the electric dipole matrix element $\vec{d}$ of the two level atom as $m \omega_0\vec{d}/e\times\hat{\sigma}_{y}$, where $\hat{\sigma}_{y}:=i(\hat{\sigma}_{12}-\hat{\sigma}_{21})$. $\hat{A}$ is the potential vector. We decompose it over a complete mode basis $\vec{e}_{\mu}$ as
\begin{eqnarray}
\hat{A}(\vec{r},t)&=& \sum_\mu~ \sqrt{\frac{2\pi \hbar c^2}{\omega_\mu}}~\vec{e}_{\mu}(\vec{r})~\hat{a}_{\mu}(t)+h.c.,
\end{eqnarray}
where the sum is to be taken over {\it all normalized} eigenfunctions $\vec{e}_{\mu}$ of the Helmholtz equation.

%

In the Markov regime, that is when the reservoir and atom are correlated within a short time only, one {\it can define} spontaneous emission rate and level shifts and, after solving the Heisenberg equation, get to
\begin{eqnarray}\nonumber
\frac{\partial}{\partial t} \langle \hat{\sigma}(t)\rangle&=&-\big[\gamma(\vec{r})/2+i(\Delta_e(\vec{r})-\Delta_g(\vec{r}))\big]\langle\hat{\sigma}(t)\rangle.
\end{eqnarray}
$\langle.\rangle$ denotes the expectation value over a separable initial atom/vacuum-field state.
We write
\begin{eqnarray}\label{dec1}
\gamma(\vec{r})&=&2\sum_\mu |g_{\mu}(\vec{r})|^2 \delta(\omega_\mu-\omega_0),\\ \label{S1}
\Delta_{g,e}(\vec{r})&=&\sum_\mu |g_{\mu}(\vec{r})|^2 P\Big[\frac{1}{\omega_\mu\pm\omega_0}\Big],
\end{eqnarray}
where the $+/-$ hold for the ground/excited state shift respectively, and
\begin{eqnarray}
g_{\mu}(\vec{r})=\omega_0\sqrt{\frac{2\pi}{\hbar \omega_\mu}} \big[ \vec{d}\cdot \vec{e}_{\mu}(\vec{r})\big],
\end{eqnarray}
is the vacuum Rabi frequency of the mode $\mu$.
The sum over the $\mu$ eigenmodes runs up to the Bethe momentum cut-off, $K=mc/\hbar$ \cite{Mil94}.

The excited state shift is due to the emission of real photons, whereas the ground state level shift arises from absorption and emission of virtual photons from the vacuum reservoir \cite{Mil94}. To obtain the correct Lamb shift in the non-relativistic theory however, we have to add the term proportional to $\hat{A}^2$ in the interaction Hamiltonian, which was actually discarded when getting to Eq.~(\ref{S1}).  We come back to this in section \ref{VIR}, where we calculate the Lamb shift in the three-dimensional case. 

We would like to emphasize that as we only consider the atom to be far from the mirror ($k_0 R\gg 1$, where $k_0=\omega_0/c$), the modification of the vacuum {\it mode density}, affects the atom only if it is in the ground state \cite{Hin91}. The change in the excited state properties are here due to a pure self-{\it interference} of the electromagnetic modes that the excited atom can couple to \cite{Hin91,Mil94}.
We will then neglect the dependence of the coupling strength on $\omega_k$ for the excited level shift, which comes down to ignoring energy level shifts (Van der Waals shifts) that are significant only if the mirror
is very close to the atom. For a two level atom, these shifts
are identical for the excited and ground state \cite{Hin91}, so the total 'near field'
energy shift would remain unaffected by the presence of the mirror anyway.

Before moving to the three-dimensional results, and to gain physical insight on the process of emission and energy shifts in the presence of a mirror,
we first perform a one-dimensional calculation.\\

\section{One-dimensional Model}
In this section, we here assume that the atom can only couple to the electromagnetic fields through a set of one-dimensional spatial modes $k$ along the mirror axis $z$ This means that the other modes do not contribute to spontaneous emission.
This situation of course bears a similarity to the spherical mirror case, where also half of the light field is reflected back to the atom, and is therefore worth investigating in some details.

\subsection{Normal mode and quantization}

The scalar, one-dimensional, mode functions ${e}_{k}(z)$ of the problem must satisfy the Helmholtz equation
\begin{eqnarray}
\nabla^2~{{e}}_{k}(z)+|k|^2~{{e}_{k}}(z)=0.
\end{eqnarray}
The solutions are superpositions of plane waves traveling in reverse directions.
We assume that the (one-dimensional) quantization domain has a length $L$, that the atom is at $z=0$, and the mirror at $z=-R$.
The density of $k$ modes is readily found to be $L/2\pi$ from the periodic boundary conditions at $z=L$.
From the boundary condition on the mirror ${e}_{{k}}(z=-R)=0$, and after normalizing each mode to unity, we find that the mode functions can be written
\begin{eqnarray}
{e}_{{k}}(z)=\frac{1}{\sqrt{2L\mathcal{A}}}(e^{ikz}-e^{-2ikR}e^{-ikz}),
\end{eqnarray}
where $\mathcal{A}$ is the transversal cross section area of the field around the atom.
It is clear from this relation, that the wave going to the mirror (second term), and the wave going directly to free space (first term) will interfere perfectly provided temporal coherence is fulfilled.

Using this mode function, we now compute the influence of the mirror on an excited atom, i.e on {\it real photon} processes .

\subsection{Real photon processes, atom in the excited state}

We here neglect the modification of the mode spectral density that couples to the excited atom, and consider the modification of the atomic state due to self-interference, as already discussed, so we neglect the dependence of $g_\mu$ on $\omega_\mu$.
In the Markov regime, using Eq.~(\ref{dec1}) and (\ref{S1}), we find that the atomic coherence decay and excited state level shift at $R=0$ are then
\begin{eqnarray}
\gamma(0)&=&\gamma_{\rm FS}(1-\cos(2k_0R))\\
\Delta_e(0)&=&\gamma_{\rm FS}\sin(2k_0R),
\end{eqnarray}
where the free space 1D spontaneous emission rate is
\begin{eqnarray}
\gamma_{\rm FS}=\frac{2d^2\omega_0}{\hbar \mathcal{A}}.
\end{eqnarray}
For $2k_0R=2\pi n$, ($n$ an integer number), we get a complete suppression of spontaneous emission and excited level shift. On the other hand, for
$2k_0R=\pi n$, the spontaneous emission is enhanced by a factor of two.

The reason for such large effects is that the fields going to the mirror and the `direct' fields can fully interfere in the Markov regime.
To find out in which regime temporal coherence is not-satisfied, so that the visibility of the single photon interference is not perfect, we would need to consider the exact quantum dynamical evolution without making a Markov approximation. Such an analysis would reveal that temporal coherence is reduced when the mirror is placed such that the time it takes for the light field to go to the mirror and back is larger than
the atom decay time. This scenario was investigated, for example in \cite{Dorner}, and studied experimentally in \cite{Dub07}, where signatures of non-Markovian dynamics were analyzed using a Hanbury Brown and Twiss set-up. It was noted that bunching appears for short time scales, similar than the bunching that would appear for two classical sources (which here would be the atom and its far-distant mirror image).

In the `extreme' non-Markovian regime, where the mirror is placed far away from the atom, a multimode field with width $1/\gamma_{\rm FS}$ is emitted towards the mirror with 50\% probability. The other half goes to free space. The atom is completely in the ground state when this field returns from the mirror. It will re-excite the atom after a time $\tau=2R/c$, but only partially since its temporal envelope is not the time-reversed spontaneously emitted field \cite{Sto09} and that its amplitude is twice as small.
After such a (partial) re-excitation, another field will be emitted along the mirror so that the atom will again be re-exited later on. Eventually, the photon will leave after a complex dynamical process that resembles that of a multi-mode cavity \cite{Dorner,Ris08}.
In this paper, we always assume a Markovian dynamics, where a linear decay and level-shifts can be defined. We will only be concerned with spatial decoherence, by assuming ideal temporal overlap of the single photon with itself, i.e $\tau \ll 1/\gamma_{\rm FS}$.

We assumed here that the atom only couples to the longitudinal modes along the mirror axis, which is not realistic for an atom in free space. We will now calculate the effects of polarization and use a spherical mirror. We will show that similar behavior appears for a full hemisphere, i.e in the Markovian regime, spontaneous emission and excited state level shifts can be suppressed. We also calculate the far field Casimir-Polder shift and compare it with the well-known calculations of the Casimir-Polder shift for an atom close to an infinite plane mirror.

\section{Three-dimensional Model}\label{sec2}

In this section, the space will be divided into a part inside a sphere of radius R (region I), and the annular region between the sphere of radius R and the quantization sphere (region II). See Fig. \ref{notations}.

\subsection{Normal modes}

In spherical coordinates, $\mu=(l,m,\sigma)$ where $\sigma$ denotes the TE or TM modes, and $(l,m)$ the quantum numbers for the angular momentum and spin respectively.
The solution of the Maxwell equation for the electric field can be written as a superposition of the electric and magnetic multipoles \cite{Jac}
\begin{eqnarray}
\vec{e}_{\rm{TM}}(\vec{r})&=&g_l(k_\mu r)\vec{X}_{l,m}(\Vec{\Omega})\\
\vec{e}_{\rm{TE}}(\vec{r})&=&\frac{i}{k_\mu}\vec{\nabla}\times ( f_l(k_\mu r)\vec{X}_{l,m}(\Vec{\Omega})),
\end{eqnarray}
where $\vec{X}_{l,m}(\Vec{\Omega})=\vec{L}Y_{l,m}(\Vec{\Omega})/\sqrt{l(l+1)}$ is the vectorial spherical harmonic and
$f_l,g_l$ are superpositions of spherical Bessel or Hankel functions. $\Vec{\Omega}$ is the vectorial solid angle along the radial direction $\vec{r}$.
 $\vec{L}=\vec{r}\times\vec{p}/i$ is the angular momentum operator.
 The magnetic induction $\vec{B}$ is a superposition of the two multipoles
\begin{eqnarray}
\vec{b}_{\rm{TM}}(\vec{r})&=&\frac{-i}{k_\mu}\vec{\nabla}\times ( g_l(k_\mu r)\vec{X}_{l,m}(\vec{\Omega})),\\
\vec{b}_{\rm{TE}}(\vec{r})&=&f_l(k_\mu r)\vec{X}_{l,m}(\vec{\Omega}).
\end{eqnarray}
The radial functions are written as
\begin{eqnarray}\label{RADIALMODES}
g_l(k_\mu r)&=&c_l j_l(k_\mu r)~~{\rm in~region~I},\\
&=&a_l h^{(1)}_l(k_\mu r)+b_l j_l(k_\mu r)~~{\rm in~region~II},
\end{eqnarray}
where $h^{(1)}_l(k_\mu r)$ is the spherical Hankel function, and $j_l(k_\mu r)$ the spherical Bessel function of the first kind.
It is clear that, in the absence of the mirror, the $b_l$ modes are the vacuum field from region II. We will show next that, in the single mirror geometry, we only need to quantize these modes.

\subsection{Quantization}

We will not solve the full quantum mechanical problem and quantize the electromagnetic field in the presence of the spherical mirror. This can be done exactly in the case of a full sphere \cite{Boy68} by solving the eigenvalue equation derived from the boundary conditions. To the best of our knowledge, such a treatment has not been done for an opened geometry such as a hemispherical mirror. As was shown in \cite{Daul}, the problem is not so difficult however, if one assumes that the boundary condition lies in the far field ($k_0 R\gg 1$), as we assume here, so that the mode density is close to that of free space. We are then mostly dealing with a continuum of modes, like in the 1D calculations.

Using the boundary condition on the large quantization sphere, we find that
\begin{eqnarray}
k_l \Lambda=l\frac{\pi}{2}.
\end{eqnarray}
The density of free space vacuum modes is then $2\Lambda/\pi$.
From a point close to the focal point, the $b_l$ modes are non-degenerate if the mirror is large enough, so they are all orthogonal to each other.
We normalize them so that each of them contains one photon. For the magnetic multipole we then require
\begin{eqnarray}
\int_0^\Lambda r^2 dr |g_l(kr)|^2 \int_{4\pi} d\vec{\Omega} |\vec{X}_{l,m}(\Omega)|^2=1.
\end{eqnarray}
In free space, $a_l=0$, we thus get the condition
\begin{eqnarray}
|b_l|^2\approx\frac{k^2}{2\Lambda},
\end{eqnarray}
where we used the fact that the main contribution to the vacuum fluctuations stems from the far field. The same relation holds for the vacuum modes of the electric multipole.

\subsection{Free-space decay}\label{FSsection}

Having normalised the vacuum modes $b_l$, and the normal modes of the system, we can calculate the distribution of vacuum fluctuations and atomic properties in region I, and associate the eigenmodes $b_l$ to the continuous set $\vec{k}$, so that $\sum_\mu\rightarrow \Lambda/\pi\int dk\sum_{l,m}$.
For a dipole oriented along $\vec{r}$ for example, we can check that we get the usual free space spontaneous decay. Using Eq.~(\ref{dec1}), the formula for the field component along the radial direction (see Appendix), and setting $a_l$ to zero in Eq.(\ref{RADIALMODES}), we indeed find
\begin{eqnarray}
\gamma_{\rm FS}&=&\frac{\Lambda}{\pi}\int dk \sum_{l,m}|b_l|^2 l(l+1)\frac{j_l^2(kr)}{(kr)^2} |Y_{l,m}|^2 \\
&\times&\frac{2\pi \omega_0^2 d^2}{\hbar\omega_k}\delta(\omega_k-\omega_0)\\
&=&\frac{d^2\omega_0^3}{3\pi\hbar c^3},
\end{eqnarray}
which is the standard spontaneous decay rate of an atom in free space.
In the last step, we used the addition formula for spherical harmonics $\sum_m |Y_{l,m}|^2=(2l+1)/4\pi$ and the addition formula
\begin{eqnarray}\label{bla}
\sum_{l=1} l(l+1)(2l+1)\frac{j_l^2(kr)}{(kr)^2}=\frac{2}{3}.
\end{eqnarray}
One can show that the spontaneous decay rate is the same for a tangential dipole orientation.

In the next sections, we will see how the mirror imposes a fixed phase relation between the even and odd $l$ modes appearing in the sum Eq.~(\ref{bla}) used in the spontaneous decay  calculation, and how this modifies it.
This is already hinted by noting that
\begin{eqnarray}\nonumber
\sum_{l~\rm{even/odd}}l(l+1)(2l+1)\frac{j_l^2(kr)}{(kr)^2}&=&\\\label{split}
\frac{2}{3}\int_0^{\pi/2} d\theta\sin^3\theta \Big[\begin{matrix} \sin^2(kr\cos\theta)\\\cos^2(kr\cos\theta) \end{matrix}\Big],\\\nonumber
\end{eqnarray}
where the odd/even modes correspond to the sine/cosine functions.
Depending on whether the atom is at the node or the antinode of the standing wave, it couples preferentially to the even or odd modes.
Without a defined phase relation between even and odd $l$ modes, as is the case for free space, their spectral densities always adds up to 2/3 as we just saw, but they can otherwise cancel or add up coherently.
This will allow suppression or enhancement of the even or odd vacuum modes fluctuations, and thus significant changes in the decay and shifts.

\subsection{Boundary conditions}

The boundary conditions on the electric and magnetic fields are
\begin{eqnarray}\label{FSE}
\vec{n}\times\vec{E}^{\rm I}|_{r=R}&=&\vec{n}\times\vec{E}^{\rm II}|_{r=R},\\\label{FSB}
\vec{n}\times\vec{B}^{\rm I}|_{r=R}&=&\vec{n}\times\vec{B}^{\rm II}|_{r=R},
\end{eqnarray}
for $\theta=[\alpha,\pi[$ and $\phi=[0,2\pi[$ (see Fig. 1), where $\vec{n}$ is the normal to the mirror. We write $\vec{E}^{\rm I,II}$, the electric field in region I,II.
Assuming that the mirror is a perfect conductor, as we always do in the paper, we also have
\begin{eqnarray}\label{mirrorE}
\vec{n}\times\vec{E}^{\rm I}|_{r=R}&=&0,\\\label{mirrorB}
\vec{n}\cdot\vec{B}^{\rm I}|_{r=R}&=&0,
\end{eqnarray}
for $\theta=[0,\alpha[$ and $\phi=[0,2\pi]$.

Two sets of equations for the transverse electric and magnetic multipoles can then be obtained and solved for $c_l$ to calculate the total field in region I.

\subsection{System of equations}

We use the symmetry along $\phi$ by projecting the boundary conditions over the $m$ modes (multiplication by $e^{im\phi}$ and integration over $\phi$). From equation (\ref{mirrorE}), we get, for example for the $\theta$ component of TM mode, the relation
\begin{eqnarray}
\sum_{l=1}^\infty g^{I}_l(kR)A_{l,m}(\theta)&=&0; ~~{\rm for}~~\theta=[0,\alpha[,
\end{eqnarray}
where
\begin{eqnarray}
A_{l,m}(\theta)=\sqrt{\frac{(2l+1)(l-m)!}{(l+m)!}}\frac{P_l^m(\cos\theta)}{\sqrt{l(l+1)}}
\end{eqnarray}
and $P_l^m$ is the associated Legendre polynomial.
After projecting over $m$ and $l$, and using the orthogonality of the spherical modes, we require
\begin{eqnarray}
c_l j_l(kR)=a_l h_l^{(1)}(kR)+b_l j_l(kR).
\end{eqnarray}
Last, we use Eq. (\ref{FSB}) and obtain
\begin{eqnarray}
\sum_{l=1}^\infty \frac{\partial [r g^{I}_l(kr)]}{\partial r} A_{l,m}(\theta)&=&\sum_{l=1}^\infty \frac{\partial [r g^{II}_l(kr)]}{\partial r} A_{l,m}(\theta),
\end{eqnarray}
for $\theta=[\alpha,\pi]$, and $r=R$.
Using the Wronskian for spherical Bessel functions, we then get the two sets of equations
\begin{eqnarray}\label{DE}
\sum_{l=1}^\infty (c_l-b_l)\frac{A_{l,m}}{h_l^{(1)}(kR)}&=&0;~~\theta=[0,\alpha[\\\label{DE2}
\sum_{l=1}^\infty c_l j_l(kR) A_{l,m}&=&0;~~\theta=[\alpha,\pi]
\end{eqnarray}
for the coefficients of the magnetic multipole. Similarly, we get
\begin{eqnarray}\label{TEM2}
\sum_{l=1}^\infty (d_l-e_l)\frac{A'_{l,m}}{[r h_l^{(1)}(kr)]'}&=&0; ~~\theta=[0,\alpha[\\
\sum_{l=1}^\infty d_l [r j_l(kr)]' A'_{l,m}&=&0; ~~\theta=[\alpha,\pi]\label{TEM25}
\end{eqnarray}
for the electric multipole, where $[r j_l(kr)]'=\partial (r j_l(kr))/\partial r | r=R$, and $A'_{l,m}=\partial A_{l,m}/\partial \theta$.
$e_l$ and $d_l$ are the amplitude coefficients for $f_l(kr)$ equivalent to $c_l$ and $b_l$ for $g_l(kr)$.
Each set of equations can be written as a Fredholm equation that can be solved numerically \cite{Col62}.

Here, we will show that an analytical solution can be found in the limit
where $k_0R\gg l(l+1)$, that is using a large mirror and/or looking at the field fluctuations close to the focus.
We note that the field is orthogonal to $\vec{n}$ far from the origin (See Eq.~(\ref{dec})). The condition (\ref{mirrorB}) is then satisfied automatically. In this case, the solutions of the two sets of equations will be identical for both the TE and TM modes, so that the reflection off the mirror will preserve polarisation.

The solution can then be found using scalar fields. We will include the polarisation dependence of the dipole emission later, after having identified far field plane wave modes. We will here write
\begin{eqnarray}\label{scal}
\phi_b(\vec{r})&=&\sum_{l,m} c_l~j_l(kr) Y_{l,m},
\end{eqnarray}
for the total field amplitude in region I, and solve for $c_l$ as a function of $b_l$ using the two equations
\begin{eqnarray}\label{DE3}
\sum_{l=0}^\infty (c_l-b_l)\frac{Y_{l,m}}{h_l^{(1)}(kR)}&=&0; ~~\theta=[0,\alpha[\\\label{DE4}
\sum_{l=0}^\infty c_l j_l(kR) Y_{l,m}&=&0;~~\theta=[\alpha,\pi].
\end{eqnarray}
We simply removed the $l(l+1)$ dependance of the field modes, which as can be found from Eq.~(\ref{split}), is equivalent to ignoring the polarization dependance of the dipole emission ($\sin^2\theta$ in the integral).
Note that the summation starts at $l=0$ now, as is allowed for scalar fields. This is not true for the general solution of the (vectorial) Maxwell equation which does not have spherically symmetric solutions.

\subsection{Full-hemispherical mirror}

We first assume that the mirror covers $2\pi$ steradian, so $\alpha=\pi/2$.
Furthermore, since $k_0 R \gg l(l+1)$ we expand the Bessel functions in the far field.
\begin{eqnarray}
h_l^{(1)}(kR)&\approx&(-i)^{l+1}e^{ikR}/kR,\\
j_l(kR)&\approx&\sin(kR-l\pi/2)/kR.
\end{eqnarray}
Using the relation
\begin{eqnarray}
\int_0^1 P_l^m(x) P_{l'}^m(x) (1+(-1)^{l+l'})=2\frac{\delta_{l,l'}}{2l+1}\frac{(l+m)!}{(l-m)!},
\end{eqnarray}
in Eq.~(\ref{DE3}) and Eq.~(\ref{DE4}) and then solving for $c_l$, we obtain after some algebra, the two relations
\begin{eqnarray}\label{FM}
c_{l'}&=&e^{ikR}\cos(kR)\big( b_{l'}-i\sum_{m,o}b_l I_{l',l,m} \big)~\rm{for}~\it{l'}~\rm{even},\nonumber\\
c_{l'}&=&i e^{ikR}\sin(kR)\big( b_{l'}+i\sum_{m,e}b_l I_{l',l,m} \big)~\rm{for}~\it{l'}~\rm{odd},
\end{eqnarray}
where
\begin{eqnarray}
I_{l',l,m}=(-1)^{(l+l'+1)/2}\int_\phi\int_{\theta=[0,\pi/2]} Y_{l,m}Y_{l',m}d\Vec{\Omega}.
\end{eqnarray}
We denoted $\sum_{o,e}$ the sum over odd/even $l$ modes.
The two terms on the right hand side of each equation are the reflected and incoming vacuum amplitudes contributions to the field in region I.

At the focus, only the $l'=0$ mode is predominant since the radial amplitudes (given by $j_{l'}(kr)$) are negligible for higher $l'$.
The total field $\phi_b(\vec{r})$ at the center will then be zero for $kR=n\pi$, as can be seen from Eq.~(\ref{FM}) and (\ref{scal}).
If we draw any standing wave from the mirror through the origin to the other free space boundary with the condition that $kR=n\pi$,
it is then invariant under rotation upon $\theta$. It follows that to completely describe the field in region I
when the total field is zero at the focus, we only need even modes (as is apparent from Eq.~(\ref{FM})).
For the case where $kR=n\pi/2$, we will however need the odd modes to find the total field in region I.
This two extremal conditions explain why even and odd $l'$ modes behave differently with respect to mirror positions, as we already anticipated in section (\ref{FSsection}).

We also note that, contrary to what one would find for a mirror covering $4\pi$ (which would behave as a cavity), the angular asymmetry of the hemisphere does not yield a one to one mapping of the free space modes $b_l'$ of region II,  to the $c_l'$ modes of region I. The total far field amplitude that can enter region I is here necessarily a superposition of even and odd modes.

We already showed that the total field at the focal point $\phi_b(\vec{r}=0)$, will be zero at a node. Spontaneous emission and level shifts will then certainly cancel. To demonstrate this result, and also treat the case of a finite mirror size, it is useful to introduce some relations between the plane waves and spherical harmonics.

\subsection{Plane wave decomposition}\label{PWD}

Let us first write the far field amplitude in region I as
\begin{eqnarray}
\sum_{l,m} b_l j_l(kr)Y_{l,m}&=&i f^o(\vec{\Omega})\frac{\rm cos(kr)}{kr}+f^e(\vec{\Omega})\frac{\rm sin(kr)}{kr},
\end{eqnarray}
where
\begin{eqnarray}
f^{e/o}(\vec{\Omega}):=\sum_{m,e/o} b_l~i^l Y_{l,m}(\vec{\Omega}),
\end{eqnarray}
is the scattering amplitude for the even/odd mode. It is easy to show that the superpositions
\begin{eqnarray}
f^{\rm in}(\vec{\Omega})=\frac{i}{2}(f^o-f^e)~~,~~f^{\rm out}(\vec{\Omega})=\frac{i}{2}(f^o+f^e),
\end{eqnarray}
correspond to incoming and outgoing angular amplitudes respectively, and
that they are connected via
\begin{eqnarray}\label{IO}
f^{\rm out}(\vec{\Omega})=-f^{\rm in}(-\vec{\Omega}):=\hat{P} f^{\rm in}.
\end{eqnarray}
Here $\hat{P}$ is the parity operator. This relation shows the Gouy phase shift acquired by the incoming field
as it turns after focussing into an outgoing field.
The same relations also hold for the far field amplitude $g(\vec{\Omega})$ of the field in region I, where one can define
\begin{eqnarray}\label{regionI}
g^{e/o}(\vec{\Omega}):=\sum_{m,e/o} c_l~i^l Y_{l,m}(\vec{\Omega}).
\end{eqnarray}
The total field amplitude at any point of region I can in fact be written as a superposition of plane waves weighted by the far field amplitudes $g(\vec{\Omega})$.
Using the expansion of the plane waves in spherical harmonics
\begin{eqnarray}
\sum_{m,(e/o)}i^l(2l+1)j_l(kr)Y_{l,m}=\Big[\begin{matrix}   \cos(\vec{k}\cdot\vec{r})\\  \sin(\vec{k}\cdot\vec{r}) \end{matrix}\Big],
\end{eqnarray}
and Eqs.~(\ref{regionI}) and (\ref{scal}), we indeed obtain
\begin{eqnarray}\label{PW2}
\phi_b(\vec{r})&=&\frac{1}{2\pi}\int_{2\pi} d\vec{\Omega} \big[g^e \cos(\vec{k}\cdot\vec{r})+ig^o\sin(\vec{k}\cdot\vec{r})\big],
\end{eqnarray}
where we wrote $\int_{2\pi}:=\int_\phi\int_{\theta=[0,\pi/2]}$ for simplicity.
The field inside region I, is uniquely given by the coefficients $g^{e/o}$, which are linked to the incoming vacuum modes $f^{\rm in}$ via the boundary conditions.
This treatment was also used in \cite{Daul} for an open cavity.

\subsection{Finite size mirror}

To treat the case of finite mirror size, we decompose the boundary conditions into three interfaces: the mirror ($\theta=[0,\alpha[$), the opposite free space interface between region I and II ($\theta=[\pi-\alpha,\pi]$), and the remaining free space ($\theta=[\alpha,\pi-\alpha[$). As before, we project the equations Eq.~(\ref{DE3},\ref{DE4}) over the $m$ modes and using the results of section (\ref{PWD}) get to the results
\begin{eqnarray}\label{first}
\big(1-e^{2ikR}T(\theta)\big)g^o&=&-2e^{ikR}\sin(kR)\big(f^{\rm in}+T(\theta)f^{\rm out}\big),\nonumber\\
\big(1+e^{2ikR}T(\theta)\big)g^e&=&-2ie^{ikR}\cos(kR)\big(f^{\rm in}-T(\theta)f^{\rm out}\big),\nonumber\\
\end{eqnarray}
where $T(\theta)$ is a function that is zero for $\theta=[0,\alpha[$, and unity for $\theta=[\alpha,\pi/2[$. One can check that setting $T(\theta)$ to zero for $\theta=[0,\pi/2]$, and going back to the spherical basis, we recover Eq.~(\ref{FM}).

Using Eq.~(\ref{IO}), and the fact that $T$ is only zero or one, we can rewrite Eq.~(\ref{first}) as
\begin{eqnarray}
g^o&=&i\Big[T(\theta)(1-\hat{P})+2i(1-T(\theta))e^{ikR}\sin(kR)\Big]f^{\rm in},\nonumber\\
g^e&=&i\Big[T(\theta)(1+\hat{P})-2(1-T(\theta))e^{ikR}\cos(kR)\Big]f^{\rm in},\nonumber\\
\end{eqnarray}
which uniquely relates the incoming field from region II to the far field modes in region I.
With the use of the relation (\ref{PW2}), we finally get to the result
\begin{eqnarray}
\phi_b(\vec{r})=\sum_{l,m} b_l i^l T_{l,m},
\end{eqnarray}
where
\begin{eqnarray}\label{ModeFunc}
T_{l,m}&=&2i\int_{2\pi} \frac{d\vec{\Omega}}{4\pi} \Big[T(\theta)(e^{i\vec{k}\cdot \vec{r}}+ e^{-i\vec{k}\cdot \vec{r}} \hat{P})\nonumber\\
&+&2(1-T(\theta))e^{ikR}\cos(kR+\vec{k}\cdot \vec{r})\Big]Y_{l,m}.
\end{eqnarray}

The total field in I is a superposition of waves that travel through the point of coordinate $\vec{r}$ without being reflected by the mirror (first term in Eq.~\ref{ModeFunc}) and waves that are reflected by the mirror (second term) as one could have easily figured out.
A similar result was found for an open cavity with variable reflectivity in \cite{Daul}, where the density of modes of a single mirror was calculated by setting one of the mirror reflectivities to zero.
Here we derived the normal modes amplitudes for the case of a single mirror by using a Neumann condition between regions I and II  (through continuity of the $\vec{B}$ field).

This mode function can now be used to calculate the density of fluctuations around the focal point, to obtain the change in excited level shifts and spontaneous emission rates, and find the ground state Casimir-Polder shifts as a function of the mirror's numerical aperture.

\section{QED effects close to the focus of a spherical mirror}

\begin{figure}[!h]
\includegraphics[width=8cm]{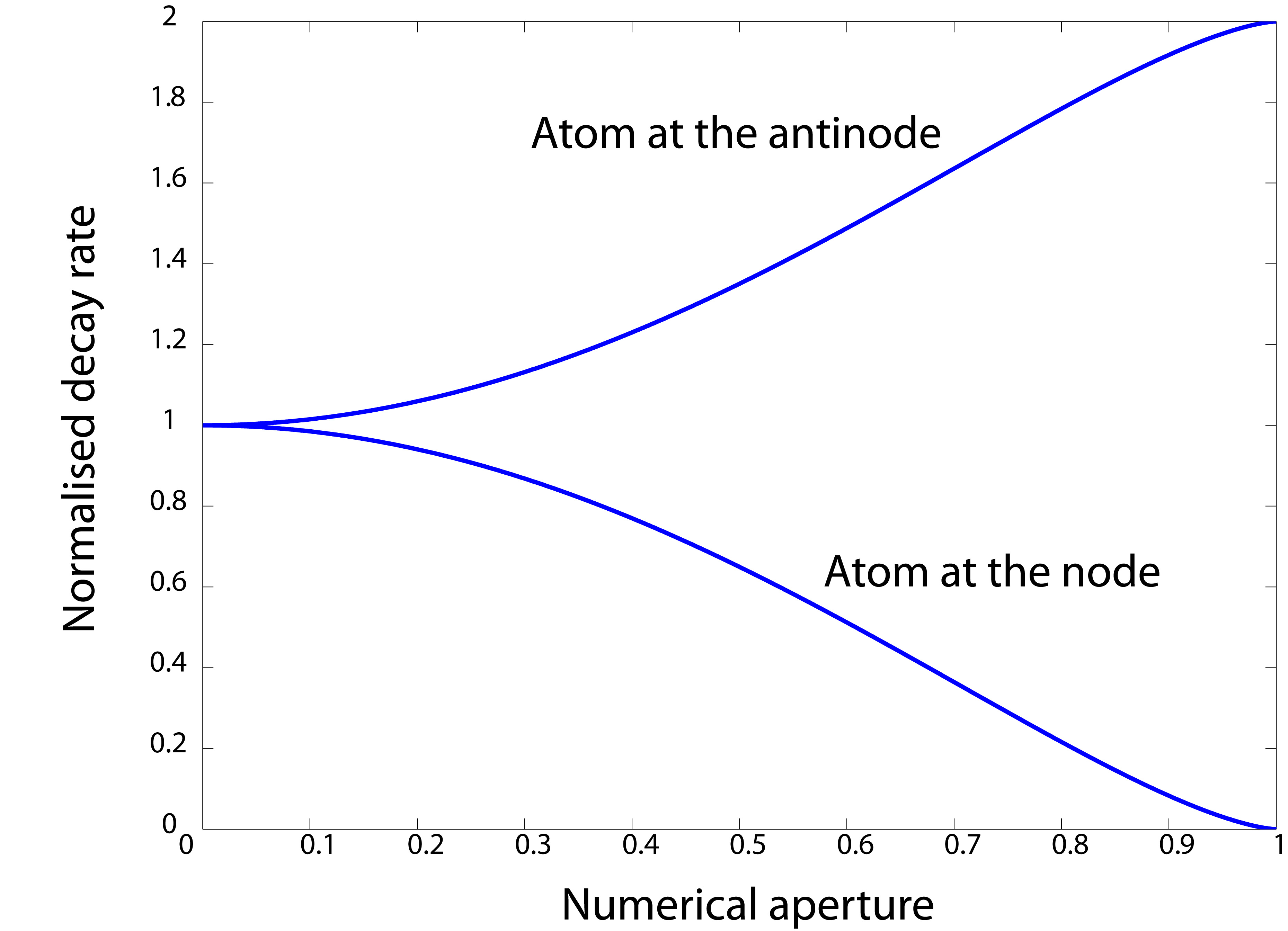}
\caption{Spontaneous decay rate as a function of the spherical mirror numerical aperture, for an atom at the mirror focus. Top curve : The atom is at the anti-node of the standing wave formed by its retro-reflected field. Bottom curve : Atom at the node of the standing wave. Large changes in the spontaneous emission rate are already expected for moderate numerical apertures. }
\label{NA}
\end{figure}

We can now calculate the QED effects on the atomic electron from Eqs.~(\ref{dec1}), (\ref{S1}) and (\ref{ModeFunc}).
The modified decay and level shifts are given by
\begin{eqnarray}
{\gamma}(\vec{r})=\frac{2 d^2\omega_0 \Lambda}{\hbar c}|\phi_{\vec{k}_0}(\vec{r})|^2,
\end{eqnarray}
and
\begin{eqnarray}
\Delta_{e,g}(\vec{r})=\frac{2 d^2 \Lambda \omega_0^2}{\hbar c} \int \frac{dk}{k} |\phi_{\vec{k}}(\vec{r})|^2 P\Big[\frac{1}{k\pm k_0}\Big],
\end{eqnarray}
%
where
\begin{eqnarray}
|\phi_{\vec{k}}(\vec{r})|^2=\frac{k^2}{ \Lambda}\int_{2\pi} \frac{d\vec{\Omega}}{4\pi}(1-\rho(\Vec{\Omega})\cos(2(k R+\vec{k}\cdot \vec{r})).
\end{eqnarray}
The dependence of the function $\phi$ on the $b$ modes is now explicitly given by $\vec{k}$.
In the last equation, we used the fact that the spherical harmonics form an orthonormal set of modes. We also introduced $\rho=1-T$, the reflectivity of the mirror.

We can now also include polarisation, and get
\begin{eqnarray}
|\phi_{\vec{k}}(\vec{r})|^2&=&\frac{k^2}{ \Lambda} \int_{2\pi} \frac{d\vec{\Omega}}{4\pi}\frac{3}{2}\Big[1-\Big|\frac{\vec{d}\cdot \vec{\Omega} }{d}\Big|^2\Big]\nonumber\\
 &\times&\Big[1-\rho\cos(2(k R+\vec{k}\cdot \vec{r}))\Big].
\end{eqnarray}
If we set $\rho$ to zero everywhere, we recover the density of vacuum fluctuations in free space.

\subsection{Real photon processes}

The spontaneous decay, normalized to the free space decay rate $\gamma_{\rm FS}$, is
\begin{eqnarray}
\overline{\gamma}(\vec{r})&=&\frac{3}{2}\int_{2\pi}  \frac{d\vec{\Omega}}{4\pi}\Big[1-\Big|\frac{\vec{d}\cdot \vec{\Omega} }{d}\Big|^2\Big] \nonumber\\
 &\times& \Big[1-\rho\cos(2(k_0 R+\vec{k_0}\cdot \vec{r}))\Big].
\end{eqnarray}
This quantity is plotted in figure \ref{NA}, as a function of the numerical aperture (defined as NA$=\sin(\alpha)$), for a linearly polarized dipole positioned at $\vec{r}=0$, and orthogonal to the mirror axis. Spontaneous emission is found to vanish for a mirror position such that $\cos(2k_0R)=0$, i.e at the node.
A twofold increase in the spontaneous emission rate is found when $\cos(2k_0R)=1$, at the antinode.

We note that for the numerical aperture used in \cite{Esc01}, (NA=0.4, i.e 4$\%$ of solid angle), a spontaneous emission rate change of 24$\%$ is predicted. Such a, perhaps unexpectedly, large change of the decay rate may be understood by noting that the factor of two coming from the interference between the two reflected and direct amplitudes, translates into a factor of 4 in density (for `small' numerical apertures). Together with the inclusion of the polarisation properties of the dipole emission, another factor of $3/2$ is gained, which in total gives $(3/2\times 4)\times 0.4=0.24$.
 The difference between the observed $1\%$ change of the excited state population in \cite{Esc01} and the $24\%$ modification of the spontaneous emission predicted here can be partly explained by residual atomic motion, finite spatial overlap, finite temporal coherence or multi-level effects in the experiment.

\begin{figure}[!h]
\includegraphics[width=8cm]{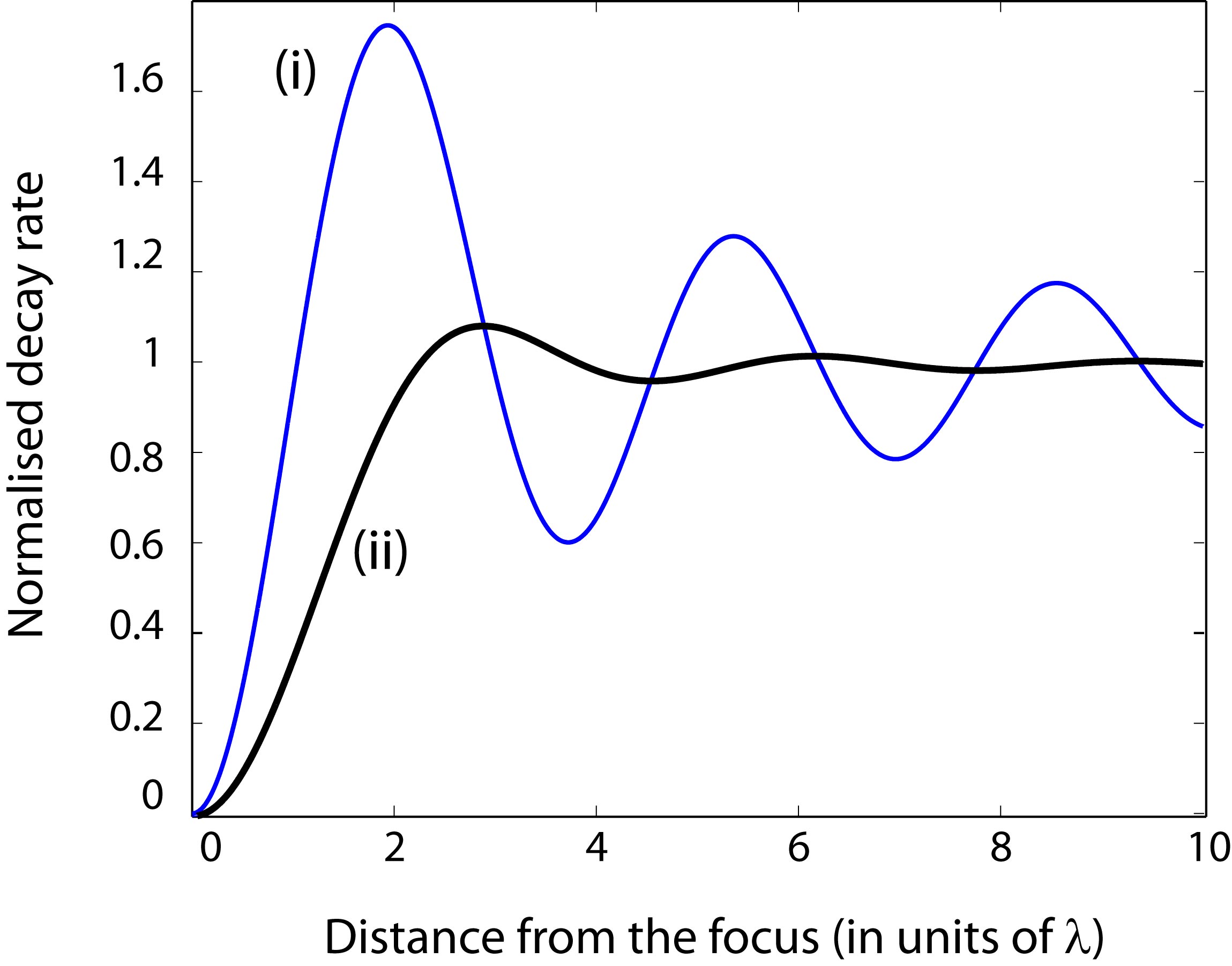}
\caption{Spontaneous decay rate of an atom as it is displaced from the focus, in the case of a full hemispherical mirror. Here the mirror is positioned such that there is a node of the standing wave at the focus. Trace (i) corresponds a scenario where the atom is displaced along mirror axis. (ii) is when the atom displaced perpendicularly to the mirror axis.} \label{pos}
\end{figure}

We can compute the modification of the decay rate as a function of distance from the focus to estimate the sensitivity to mirror, or lens, positioning. Fig.~\ref{pos} shows the dependance of  $\gamma$ for an atom that is displaced away from the focus of a hemispherical mirror and where the mirror is positioned such that $|\phi_{\vec{k}}(\vec{r}=0)|^2=0$.
The spontaneous emission rate is close to zero within a volume $\lambda^3$ around the focus, and oscillates for a few wavelengths until it reaches the free space value. More precise formulae must however be used when the atom is far from the focus, as the approximation $kR \gg l(l+1)$ is no longer valid for large distances from the focus \cite{Daul}.

The excited state level shift will also be altered in the same way due to a modified coupling to the retroreflected modes. We find after contour integration, that the excited state shift, normalized to the free space decay rate, is
\begin{eqnarray}
\overline{\Delta}_e(\vec{r})&=&\frac{3}{2}\sum_b|\phi_b(\vec{r})|^2 P\Big(\frac{1}{\omega_b-\omega_0}\Big)  \\
&=&\frac{3}{2}\int_{2\pi}  \frac{d\vec{\Omega}}{4\pi}\Big[1-\Big|\frac{\vec{d}\cdot \vec{\Omega} }{d}\Big|^2\Big] \Big[\rho\sin(2(k_0 R+\vec{k_0}\cdot \vec{r}))\Big]\nonumber
\end{eqnarray}
This gives an oscillatory level shift of amplitude $\rho$ at the focus.
For a full half-mirror, the level shift completely cancels for mirror positions such that $k_0 R=n\pi$, where the decay rate is also zero.
Its evolution with numerical aperture is similar to the spontaneous emission rate change.
Such large level shifts variations can yield a strong confining potential and would be interesting to study experimentally with large numerical apertures, similarly to what was done in \cite{Bus04}.

\subsection{Virtual photon processes}\label{VIR}

The Lamb shift of the ground state can be computed in the very same way as the excited state shift.  In the simple case of a full half-mirror, and with the atom at the mirror focus, we get
\begin{eqnarray}\label{LS1D}
\Delta_g(0)=\frac{\gamma_{\rm FS}}{k_0}\int_0^{mc/\hbar} dk \frac{k}{k_0+k}\sin^2(kR).
\end{eqnarray}
We can write this result as a sum of three terms that can be easily integrated : The electron self-energy, the free space Lamb shift, and the Casimir-Polder shift. The electron self-energy is
\begin{eqnarray}
\Delta^{se}_g=\gamma_{\rm FS}\frac{K}{2 k_0}\Big[1+\frac{\sin(2KR)}{2KR}\Big],
\end{eqnarray}
where we wrote $K=mc/\hbar$. This quantity is identified by setting $k_0$ to zero in Eq.~(\ref{LS1D}) and in fact cancels with the shift from the
$\hat{A}^2$ part of the Hamiltonian, as can be easily checked.
This procedure is known as mass renormalisation \cite{Mil94}.
The free space Lamb shift is
\begin{eqnarray}
\Delta^{fs}_g=\gamma_{\rm FS}\log\Big[\frac{k_0}{K+k_0}\Big].
\end{eqnarray}
The modified Lamb shift (or Casimir-Polder shift), the only observable quantity, is
\begin{eqnarray}
\Delta^{cp}_g=\gamma_{\rm FS}\int_{2K R}^{2k_0R} \frac{dx}{x} \cos(x-2k_0R).
\end{eqnarray}
The Casimir-Polder shift goes to zero for very large mirror-atom distances ($2k_0 R\gg 1$) as expected,
and can be approximated by
\begin{eqnarray}
\Delta^{cp}_g=\frac{\gamma_{\rm FS}}{(k_0 R)^2},
\end{eqnarray}
closer to the mirror (but always at the focus).
We note that $\Delta^{cp}_g$ drops slower with distance than in the plane mirror case, where it decreases like $\gamma_{\rm FS}/(k_0 R)^4$ \cite{Cas48}. The difference lies in the fact that there, the mirror does not cancel as many electromagnetic modes, which yields a faster decay of the QED effects with distance.

As an example, using a decay rate of 15 MHz, a wavelength of 493nm (the $S_{1/2}$ to $P_{1/2}$ transition of Barium $138$Ba$^{+}$) and a mirror distance of 1 cm, gives a Lamb shift of 100 Hz which is experimentally measurable using modern spectroscopic tools \cite{Roo06}. The complete level structure will have to be used for a precise estimation of the total shift \cite{Bar74,Hin91}, but this result is encouraging for experimental investigations of Casimir-Polder shifts using trapped atoms.

The associated force on a trapped atom due the Casimir-Polder shift may be computed easily from the above formula \cite{Daul}.
It would be interesting to study how Casimir photons created by an oscillating mirror \cite{Riz07}, and also real photons `modulated' by the mirror  \cite{Gla10}, affect the center of mass motion of a trapped atom with a high numerical aperture mirror.\\

\begin{figure}[!h]
\includegraphics[width=8cm]{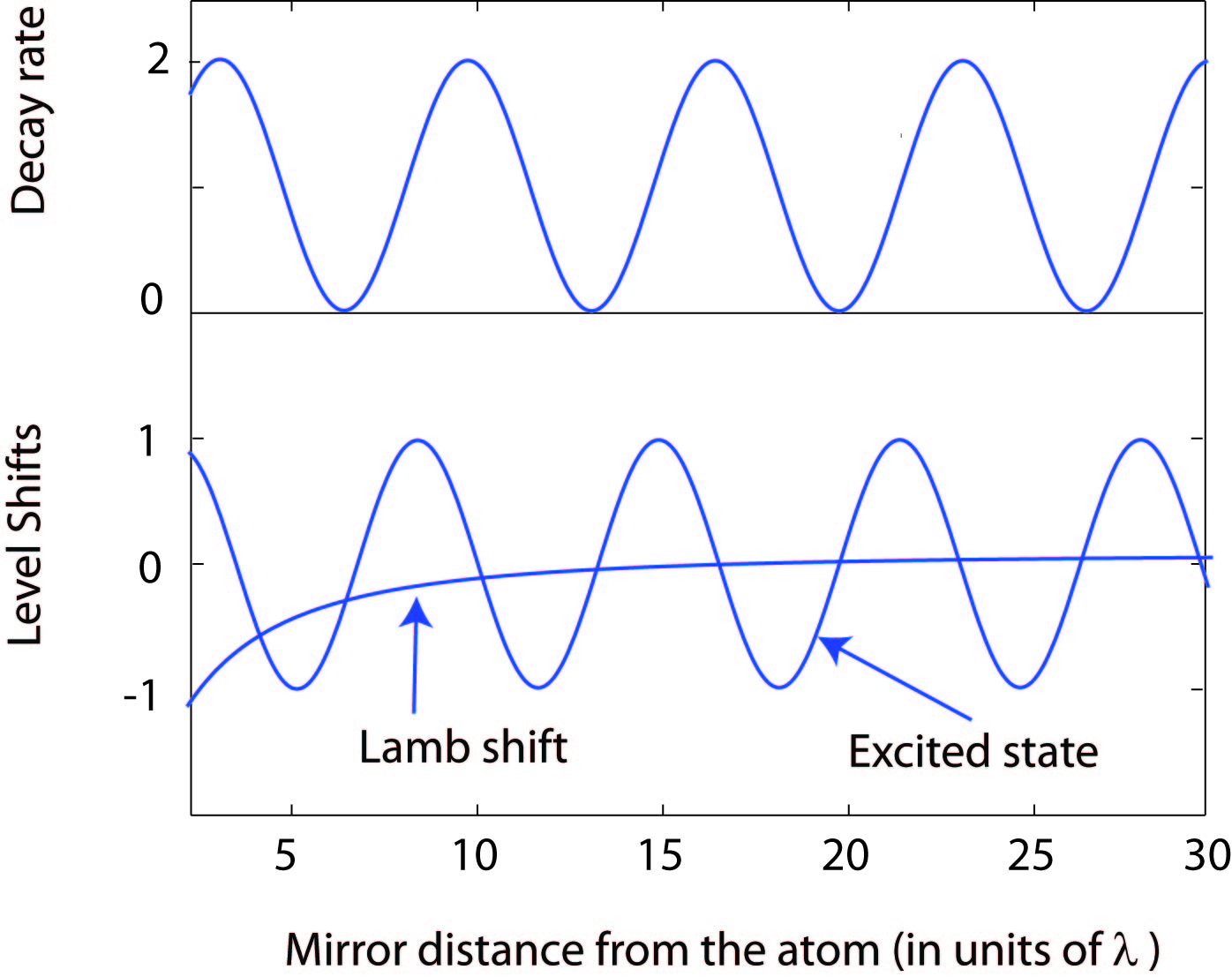}
\caption{Normalized spontaneous decay rate and excited and ground state shifts of an atom as a function of the mirror-atom distance, in the case of a full hemispherical mirror.} \label{shifts}
\end{figure}

To summarize these results, the modified decay and both ground and excited level shifts are plotted in Fig.~\ref{shifts} for a full hemi-spherical mirror.
The decay rate shows undamped oscillations between full suppression and maximum enhancement as a function of mirror distance (or mirror radius of curvature). The excited state level shift oscillates between $-\gamma$ and $\gamma$, whereas the ground state shift damps out as a function of mirror distance.

\section{Conclusion}

In conclusion, we have demonstrated that a single spherical mirror reflecting only half of the emitted field of a single atomic electron can be used
to completely suppress the atom's spontaneous emission and excited level shift.
We first presented a one-dimensional treatment that explained the underlying physics behind the full spherical mirror scenario.
The modification of QED atomic properties was then calculated as a function of the spherical mirror's numerical aperture beyond the paraxial approximation.
Large effects are found for moderate numerical apertures, and with mirror-atom distances of several wavelengths when the atom is located at the mirror focus.

This result is also relevant for the growing field of free space coupling to single absorbers, where full absorption of a single photon field requires a large coverage of the spatial dipole emission with the incoming spatial mode.
The single hemi-spherical mirror system may here serve as an efficient quantum memory that can release a stored excitation on demand on a two-level atom transition, by controlling the mirror position in a dynamical fashion.
As an application of our calculations, one also expects full super/sub radiance with two atoms interacting via large lenses covering only half of their respective dipole emission profiles, see for example \cite{Ris08,Kas05} for studies of this effect.
Last, atom trapping using the dipole force can be very efficient here, due to the steep spatial dependence of the level shift across the atom.

Finally, we calculated the Lamb shift, and showed a favorable scaling of the spherical geometry over the plane
mirror case. The Lamb shift scales as $\gamma_{\rm FS}/(k_0 R)^2$, where $R$ is the mirror radius of curvature. This contrasts with the $\gamma_{\rm FS}/(k_0 R)^4$ scaling law found for a plane mirror.
Using Rydberg atoms and high numerical aperture elements can potentially yield very large
shifts even for atom-mirror distances of a few centimeters and serve as a precise test-bed for investigations of Quantum Electro-Dynamics.

\section{Acknowledgments}

We would like to acknowledge useful discussions with C.W. Gardiner, H. Ritsch, H. Zoubi and S. Gerber, A. Daley, and P. Zoller.\\

This work has been partially supported by the Austrian Science Fund FWF (SFB FoQuS), by the European Union (ERC advanced grant CRYTERION) and by the Institut f\"ur Quanteninformation GmbH. G.~H.~acknowledges support by a Marie Curie Intra-European Fellowship of the European Union.

\section*{Appendix}

We can decompose the field in region I into radial and longitudinal parts using the relation
\begin{eqnarray}\label{dec}
\frac{1}{k}\vec{\nabla}\times (g_l(r)\vec{X}_{l,m})&=&\frac{1}{kr}\frac{\partial}{\partial r}[r g_l(r)] \vec{n}\times\vec{X}_{l,m}\nonumber\\
&+&i \frac{\sqrt{l(l+1)}}{kr}g_l(kr) Y_{l,m}\vec{n}
\end{eqnarray}
where $\vec{n}=\vec{r}/|\vec{r}|$.
Using the decomposition (\ref{dec}), and the eigenvalues of the angular momentum operator $\vec{L}$, we then get an expression
of the electric and magnetic field in spherical coordinates.

The electric field multipoles along $\phi$ read

\begin{eqnarray}
e^{\rm{\phi}}_{\rm TM}(\vec{r})&=&-ig_l(r)\frac{\partial}{\partial \theta}\frac{Y_{l,m}}{\sqrt{l(l+1)}}\\
e^{\rm{\phi}}_{\rm TE}(\vec{r})&=&\frac{im}{kr\sin\theta}\frac{\partial}{\partial r}[r f_l(r)]\frac{Y_{l,m}}{\sqrt{l(l+1)}}
\end{eqnarray}

Along $\theta$, we have
\begin{eqnarray}
e^{\rm{\theta}}_{\rm TM}(\vec{r})&=&\frac{-m}{\sin\theta}g_l(r)\frac{Y_{l,m}}{\sqrt{l(l+1)}}\\
e^{\rm{\theta}}_{\rm TE}(\vec{r})&=&+\frac{1}{kr}\frac{\partial}{\partial r}[r f_l(r)]\frac{\partial}{\partial \theta} \frac{Y_{l,m}}{\sqrt{l(l+1)}}
\end{eqnarray}
And along $r$
\begin{eqnarray}
e^{\rm{r}}_{\rm TE}(\vec{r})=-i\frac{\sqrt{l(l+1)}}{kr}f_l(r) Y_{l,m}.
\end{eqnarray}

\end{document}